\documentstyle[12pt]{article}        
\begin{document}
\title{Quantum phenomena do not violate the principle of locality\\ 
            {\large - a new interpretation with physical consequences -}} 
\author{Gyula Bene\\
Institute for Solid State Physics, E\"otv\"os University\\
M\'uzeum krt. 6-8, H-1088 Budapest, Hungary}
\date{\today}
\maketitle

\begin{abstract}
A new interpretation of nonrelativistic quantum mechanics
explains the violation of Bell's inequality 
by maintaining realism and the principle of locality.  
\end{abstract}
\section{Introduction}
A great majority of physicists seems to believe in telepathy\cite{f1}
(cf. \cite{Einstein}).  Is it a joke? No. They really do  because they 
think they have a compelling argument.
It is the experimentally proven violation of Bell's inequality.
This inequality, discovered by John Bell in 1964 \cite{Bell}, 
refers to correlations between measurements performed on two
separated quantum objects, say, on two spin-half particles. 
Note that the actual
measurements have been performed on optical photons\cite{opt}, \cite{exp}, 
\cite{exp2}, \cite{exp3} and also on 
gamma-photons \cite{phot} and protons \cite{prot}.
 These
particles were previously bound together in a singlet state, 
and became separated
in a decay process prior to the measurement. After such antecedents
it is not surprising at all that the results of spin measurements done
on the two separated particles exhibit correlations. E.g., 
as the net spin is zero, if spin is measured on both particles
in the same direction, there is a strict anticorrelation.
One expects that
the correlations are due to the common past of the particles.
This idea seems to have quantitative consequences. \cite{bellegy}

Suppose the spin component $\vec{S} \cdot \vec{a}$ has been
measured on particle 1 and the spin component $\vec{S} \cdot \vec{b}$ has been
measured on particle 2. Here $\vec{a}$ and $\vec{b}$ are unit vectors.
The shorthand notation $\vec{a}\, +$ (or $\vec{a}\, -$) will mean
that the spin component $\vec{S} \cdot \vec{a}$ has been
measured with the result $a=+ 1$ (or $a=-1$) (in $\hbar/2$ units). We shall
also write simply $a$ for the situation when
the spin component $\vec{S} \cdot \vec{a}$ has been
measured and the result of the measurement is $a$ ($=+1$ or $-1$). 
Now we assume that at each measurement there exist some stable
property $\lambda$, characterising the previous bound state of the
particles, that determines (at least in a probabilistic sense) 
the outcome of the measurements. In other terms, $\lambda$ represents
the
common past of the particles. Now we assume that \hfill\break
(*) {\em there exists a joint probability distribution 
$P(a,b,\lambda)$ for the 
possible outcomes of the measurements and
the property $\lambda$.} 

Taking into account that the particles became 
separated before the measurement, the principle of
locality implies that 
$P(a,b,\lambda)$ is of the product form
\begin{eqnarray}
P(a,b,\lambda)=
P_1(a,\lambda)P_2(b,\lambda)\rho(\lambda)\;, \label{e1}
\end{eqnarray}
where e.g. $P_1(a,\lambda)$ means the conditional probability that the 
first spin measurement along the direction $\vec{a}$ 
gives the result $a$ if the quantity characterizing
the common past was $\lambda$.

Integrating over the quantity $\lambda$,
we get for the correlation between the measurements
\begin{eqnarray}
P(a,b)=\int P_1(a,\lambda)
P_2(b,\lambda)\rho(\lambda) d\lambda \;. \label{e2}
\end{eqnarray}
As there is a strict anticorrelation for $\vec b=\vec a$,
one must have $P(\vec{a}\,+,\vec{a}\,+)=0$. 
This implies that for almost all $\lambda$ where $\rho(\lambda)\ne 0$ 
either $P_1(\vec{a}\,+|\lambda)=0$ or $P_2(\vec{a}\,+|\lambda)=0$.
We also have $P(\vec a\,-,\vec a\,-)=0$ which implies
either $P_1(\vec a \,+|\lambda)=1-P_1(\vec a \,-|\lambda)=1$ 
or $P_2(\vec a\,+|\lambda)=1-P_2(\vec a\,-|\lambda)=1$. 
As we see, $P_1(\vec{a}\,+|\lambda)$ and
$P_2(\vec{a}\,+|\lambda)$ can take on only the values $0$ and $1$,
and
\begin{eqnarray} 
P_1(\vec a\,+|\lambda)+P_2(\vec a\,+|\lambda)=1\;.\label{e2a} 
\end{eqnarray}
Bell's inequality follows now as\footnotesize
\begin{eqnarray}
P(\vec a\,+,\vec b\,+)+P(\vec b\,+,\vec c\,+)
=\int d\lambda \rho(\lambda) \left[P_1(\vec a\,+|\lambda)
P_2(\vec b\,+|\lambda)+P_1(\vec b\,+|\lambda) P_2(\vec c\,+|\lambda)\right]\nonumber\\
\ge \int d\lambda \rho(\lambda)\left[P_1(\vec a\,+|\lambda)
P_2(\vec b\,+|\lambda)P_2(\vec c\,+|\lambda)
+P_1(\vec b\,+|\lambda) P_2(\vec c\,+|\lambda)P_1(\vec a\,+|\lambda)\right]\label{e3}\\
=\int d\lambda \rho(\lambda)P_1(\vec a\,+|\lambda)P_2(\vec c\,+|\lambda)\left[
P_2(\vec b\,+|\lambda)+P_1(\vec b\,+|\lambda)\right]\nonumber\\
=\int d\lambda \rho(\lambda)P_1(\vec a\,+|\lambda)P_2(\vec c\,+|\lambda)
=P(\vec a\,+,\vec c\,+)\;,\nonumber
\end{eqnarray}\normalsize
where Eq.(\ref{e2a}) has been used to get the last line\cite{f2}.
Bell's inequality expresses the quantitative 
consequences of the idea that the correlations stem from the common 
past of the particles, provided that the principle of locality and the 
additional, natural-looking assumption (*) is true. The whole argument
is so simple and looks so convincing, that one might think
that Eq.(\ref{e3}) should actually hold. Nevertheless, calculating
the correlation quantum mechanically, the resulting
expression,
\begin{eqnarray}
P(\vec a\,+,\vec b\,+)=\frac{1}{2}\sin^2\left(\frac{\theta_{a,b}}{2}\right)\label{e4}
\end{eqnarray}
(where $ \theta_{a,b}$ stands for 
the angle between the directions $\vec a$ and $\vec b$)
 does not satisfy Eq.(\ref{e3}) for certain angles.  Indeed, chosing e.g.
 $\theta_{a,b}=45^o$, $\theta_{b,c}=45^o$ and $\theta_{a,c}=90^o$ (i.e., all the three
 spin directions lie in the same plane), we get
 $P(\vec a\,+,\vec b\,+)=\frac{1}{4}-\frac{\sqrt{2}}{8}$,
 $P(\vec b\,+,\vec c\,+)=\frac{1}{4}-\frac{\sqrt{2}}{8}$, 
 $P(\vec a\,+,\vec c\,+)=\frac{1}{2}$, in clear contradiction with Eq.(\ref{e3}).
 
The question is now what is correct in the reality: 
the quantum mechanical prediction or Bell's
inequality. In actual measurements strict anticorrelation
is hard to realize, therefore, a generalized form of Bell's
inequality \cite{CHSH} has been tested. The result of most experiments\cite{opt}, \cite{exp}, 
\cite{exp2}, \cite{exp3} (and still many others) 
violated (the generalized) Bell's inequality and was consistent
with quantum mechanics.

In case of photons the experiments of Alain Aspect et al.\cite{exp}, \cite{exp2} 
(cf. the recent Ref. \cite{exp3}) have particularly clearly demonstrated that
the quantum mechanical prediction is correct, and Bell's
inequality is indeed violated in Nature\cite{f3}.
  
As we have to accept that Eq.(\ref{e3}) fails, at least one of the
previous assumptions must be wrong. The usual conclusion\cite{Esp} is
that it is probably the principle of locality which 
fails, i.e., separated particles
can somehow influence each other, although there is no known physical
interaction between them. Note that this influence (if exists) 
should be faster than light, as the two measurements can take
place outside of each others light cone\cite{exp2}.

This conclusion questions the validity of a basic physical
principle which is well established in all the branches of physics.
Even the equations of the most sophisticated quantum field theories respect 
the principle of locality. 
Once it is satisfied by any other physical processes, 
why would quantum measurement be the only
exception? Also, it is of worth recalling that there is no direct
experimental proof for the existence of an action-at-a-distance. 
Indeed, suppose that one prepares the initial two-particle state  
several times and performs measurements on
the first particle. The statistics of the results will be the same,
whatever happens with the second particle. 
It does not matter, whether we also do some 
measurement on the second particle, provided we 
analyze only the data of the first measurement.  
So the conclusion about the failure of the principle of locality 
stems exclusively from the investigation of {\em correlations} between the
measurements done on the first and the second particle. Here, however,
we also make the additional assumption (*) which may be in error.

Why not give up then assumption (*)? Certainly, then Bell's
inequality does not follow and we do not have to give up the principle of
locality. However, it is a widely spread belief that (*) is implied by realism, i.e.
by the principle that 
the properties (in our case $a$, $b$ and $\lambda$) exist independently of
our mind. Therefore, giving up (*) would imply that
we give up realism, that is even less acceptable than giving up
locality. 

In this paper we are going to review and discuss
 such a theory\cite{bene}
where (*) is not implied by realism. In order to show that there is
such a logical possibility, it is instructive to express realism in the form that
the properties $a$, $b$, $\lambda$ are {\em elements of the reality}.
Any reasonable definition of the joint probability $P(a,b,\lambda)$ requires that 
the properties $a$, $b$, $\lambda$ be also comparable, so they must exist together, and 
the probability is assigned to the {\em set} $\{a,b,\lambda \}$. Therefore, 
the existence of $P(a,b,\lambda)$ requires that the set $\{a,b,\lambda \}$, too, 
be an {\em element} of the reality. 
Nevertheless, realism (as expressed above) implies only that the set $\{a,b,\lambda \}$
is a {\em subset} of the reality, so there is no logical necessity that
realism would imply assumption (*). In the theory presented below
we shall see that the properties $a$, $b$, $\lambda$ can be identified
with certain quantum states. Each of them can be determined by a suitable
measurement with probability $1$, without disturbing it. This is the natural
expression of realism within the theory. Nevertheless, all the three states
cannot be measured simultaneously, as the measurement of the state playing the role
of $\lambda$ will change the correlation $P(a,b)$ between the other two states.
Hence we cannot compare the three states with each other so that $P(a,b)$ is
unchanged. Correspondingly, there is no way to define $P(a,b,\lambda)$,
while the theory fulfills the requirements of realism and 
satisfies the principle of locality.

The paper is organized as follows. In Section 2. we review the new
interpretation\cite{bene} in a pedagogical style, discuss the meaning
of the new postulates and apply the scheme to an idealized quantum
measurement. In Section 3. the EPR-Bell experiment is considered. 
It is shown that the source of the correlations is exclusively
the common past of the particles, hence there is no action-at-a-distance.
It is also pointed out, that the state representing the common past
cannot be determined simultaneously with those corresponding to the
measurement results, so $P(a,b,\lambda)$ does not exist. 
The concluding Section 4. summarizes the main results of the paper
and comments briefly the relation and new achievements of the theory discussed here,
compared to previous interpretations.

\section{The new interpretation}

There are several myths about quantum mechanics today. One of them has
been mentioned above, this is the alleged nonlocality. Another one
is the deep but unfounded conviction that quantum mechanics becomes
invalid when applied to macroscopic systems. Unfounded, because there is
no experiment which would prove any significant deviation.
 Owing to the recent remarkable
experimental development in quantum optics people
 put quantum mechanics
again and again to the test, 
and quantum mechanics passes these tests without
an exception. Nonetheless, many experts have the opinion that
there must be a 'new physics' in the macroscopic regime, i.e., 
the Schr\"odinger equation must be modified there and it has
measurable effects.

The ideas to be presented here do not want to change quantum mechanics
in this way. The present author thinks that according to all the
available experimental evidence the Schr\"odinger equation is 
valid for any closed system, even if it is macroscopic. But quantum mechanics
is not solely the Schr\"odinger equation. As has been postulated
by von Neumann\cite{von Neumann}, at measurements allegedly something
very strange happens, which cannot be described 
in terms of the Schr\"odinger equation. It is the collapse (or reduction)
of the wave function. Although for practical purposes this concept
is satisfactory, at the same time it is the source of all the 
difficulties in quantum mechanics. As a matter of fact, if we think
seriously that the Schr\"odinger equation is valid, then it cannot
happen that certain distinguished physical systems, the measuring
devices, do not obey it. Therefore, the new interpretation 
does not use the concept of the collapse at all. Instead, a 
completely new physical assumption is introduced (the fundamental
dependence of the states on quantum reference systems, see below), together
with a series of new postulates. The resulting new scheme is free from
the usual conceptual difficulties\cite{bene}: it solves Schr\"odinger's cat paradox\cite{cat}
and  the Einstein-Podolsky-Rosen (EPR) paradox\cite{EPR}. As we shall demonstrate
in the present paper as well, the new interpretation also explains the violation
of Bell's inequality by maintaining both realism and the principle of
locality. 

For some people a new interpretation is just a 'rewording' of the same physics.
They should replace the word 'interpretation' with the word 'theory' 
everywhere in this paper, as
the scheme presented here has a deep physical content. The situation is
somewhat analogous to the birth of special relativity: in 1905 the Lorentz
transformation was already well known, and people tried to explain
it with the special properties of the 'aether'. The discovery of the
 actual physical content
(the fundamental dependence of the lengths and time intervals 
on the choice of the inertia system\cite{f5})
by Einstein did not change the formulas, still it made clear 
that there is no 'aether' at all, but 
one has to do with a new physical principle 
and has to revise his basic concepts about physical reality.
Similarly, today people believe in nonlocality and 
construct theories in order to explain
the collapse of the wave function. 
The present theory (or interpretation) tells us that it is useless, as 
there is no collapse at all, but we
have to do with a new physical principle (the fundamental
dependence of the states on quantum reference systems) again and we have to revise
our old ideas about quantum mechanics and the meaning of the wave function.

\subsection{The main idea}

We shall introduce the main idea of the new interpretation step by step, 
explaining it first on a simple example. 

Let us consider an idealized measurement
of an $\hat S_z$ spin component of some spin-$\frac{1}{2}$ 
particle. Be the particle $P$ initially in the state
$
\alpha |\uparrow>+\beta|\downarrow>$,
where $|\alpha|^2+|\beta|^2=1$ and the states $|\uparrow>$ and $|\downarrow>$ are the
eigenstates of $\hat S_z$ corresponding to the eigenvalues
$\frac{\hbar}{2}$ and $-\frac{\hbar}{2}$, respectively.
The dynamics of the
measurement is given by the relations
$|\uparrow>|m_0>\; \rightarrow \; |\uparrow>|m_{\uparrow}>$ and  
$|\downarrow>|m_0>\; \rightarrow \; |\downarrow>
|m_{\downarrow}>$, where $|m_0>$, $|m_{\uparrow}>$ and $|m_{\downarrow}>$ stand for states 
of the measuring device $M$ and the arrow $\rightarrow$ symbolizes the unitary time evolution. 
The linearity of the Schr\"odinger equation
implies that the measurement process can be 
written as
\begin{eqnarray}
(\alpha |\uparrow>+\beta |\downarrow>)|m_0> \;
\rightarrow \; |\Psi>=\alpha |\uparrow>|m_{\uparrow}>
+\beta |\downarrow>|m_{\downarrow}>\quad.
 \label{u2}
\end{eqnarray}
Let us consider now the state of the measuring device $M$ after the measurement. As the
combined system $P+M$ is in an entangled state, the 
measuring device has no own wave function and may be described
by the {\em reduced density matrix}\cite{Landau} 
\begin{eqnarray}
\hat \rho_M=Tr_P\left(|\Psi><\Psi|\right)
=|m_{\uparrow}>|\alpha|^2<m_{\uparrow}|
+ |m_{\downarrow}>|\beta|^2<m_{\downarrow}|\quad,\label{u3}
\end{eqnarray}
where $Tr_P$ stands for the trace operation in the Hilbert
space of the particle $P$. Nevertheless, if we look at the measuring device,
we certainly see that either $\frac{\hbar}{2}$ or $-\frac{\hbar}{2}$ spin component
has been measured, that correspond to the states $|m_{\uparrow}> $ and $|m_{\downarrow}>$,
respectively. These are not the same as the state (\ref{u3}), which contains both
$|m_{\uparrow}> $ and $|m_{\downarrow}>$, not only one of them. At this point it is
important to note that the usual 'ignorance interpretation' (which tells that 
the meaning of Eq. (\ref{u3}) is just that either $|m_{\uparrow}> <m_{\uparrow}|$ 
or $|m_{\downarrow}><m_{\downarrow}|$ is realized) is
 incompatible with the assumed universal validity of the Schr\"odinger equation. 
Indeed, the 'ignorance interpretation' would imply at once that actually
only one term of the superposition (\ref{u2}) exist, i.e., there is a collapse
of the wave function. The statement that
the density matrix of an individual system must be
an objective and irreducible property can also be derived 
from the requirement of locality \cite{mermin}.
 
So the quantum prediction does not agree with experience. Why do we get different states?
According to orthodox quantum mechanics\cite{f6}
one may argue as follows. The reduced density matrix $\hat \rho_M$ 
has been calculated from the state $|\Psi>$ (cf. Eq.(\ref{u2}))
of the whole system $P+M$. A state is the result of a measurement
(the preparation), so we may describe $M$ by $\hat \rho_M$ if we have
gained our information about $M$ from a measurement done on $P+M$. 
On the other hand, looking at the measuring device directly
is equivalent with a measurement done directly on $M$.
In this case $M$ is described by either $|m_{\uparrow}> $ or $|m_{\downarrow}>$.
We may conclude that performing  measurements on 
different systems (each containing the system we want to decribe)
gives rise to different descriptions 
(in terms of different states).
Let us call the system which has been measured (it is $P+M$
in the first case and $M$ in the second case) the {\em quantum
reference system}. Using this terminology, we may tell that
we make a measurement on the quantum reference system $R$, thus we prepare
its state $|\psi_R>$ and using this information we calculate
the state $\hat \rho_S(R)= Tr_{R\setminus S} \left(|\psi_R><\psi_R|\right)$
of a subsystem $S$. This means that we have to take the trace 
over the Hilbert space of the system $R\setminus S$ 
which is complementer to $S$ in $R$. We shall call
$\hat \rho_S(R)$  the state of $S$ with respect to $R$. 
Obviously $\hat \rho_R(R)=|\psi_R><\psi_R|$, thus $|\psi_R>$
may be identified with the state of the system $R$ 
with respect to itself. For brevity we shall call both $\hat \rho_R(R)$ and
$|\psi_R>$ the internal state of $R$.

Let us emphasize that up to now, despite of the new terminology,
we have not made any new physical assumptions. We have merely 
considered some rather elementary consequences of basic quantum mechanics. 

Mathematically it is evident that the state of
the system $S$ (i.e., $\hat \rho_S(R)$) depends
 on the choice of the quantum reference
system $R$. Now we ask again, what is the physics behind it?

In the spirit of the Copenhagen interpretation
one would answer that in quantum mechanics measurements
unavoidably disturb the systems, therefore, if we perform
measurements on  different surroundings $R$, this disturbance is
different, and this is reflected in the $R$-dependence of $\hat \rho_S(R)$.
Nevertheless, this argument is not compelling. 

At this decisive point we leave the traditional framework of quantum mechanics
and 
assume that the states $\hat \rho_S(R_1)$, $\hat \rho_S(R_2)$... have already existed before the
measurements. It means that their difference cannot be attributed to the measurements.
Hence the $R$-dependence
of $\hat \rho_S(R)$ is a fundamental feature, which
cannot be explained. We have to accept it as a basic property of Nature. Let us emphasize 
that it is a radical departure from traditional quantum mechanics. The states are
not conceived as the results of some measurements, instead, they are elements of the reality (more will
be said about this below). Therefore, according to the present interpretation
the famous saying of J.A.Wheeler ('A quantum phenomenon is a phenomenon only
if it is a recorded phenomenon') does not hold true. Concerning the relation between states and measurements, 
measurements will be conceived as usual interactions between physical systems
and will be explained in terms of the states, not vice versa. 

Now we have the picture that a given system $S$ can be characterized at one and the same instant of time 
 by a multitude of states  $\hat \rho_S(R_1)$, $\hat \rho_S(R_2)$..., each
referring to a different quantum reference system $R_1$, $R_2$,  etc. These quantum reference systems
themselves are physical systems which contain the system $S$. We have to explain, what 
it means that these states are elements of the reality. In the present theory the elements of reality
will be defined rather similarly (but not equivalently) to the well-known EPR reality criterion\cite{EPR}. We shall
say that 

{\em a state $\hat \rho_S(R)$ is an element of the reality if and only if there exist in principle
a suitable measurement, which, if performed on the quantum reference system $R$, does not change
this state and determines it with unit probability.}

We shall call the above 'suitable measurement' a {\em nondisturbing measurement}.
The term {\em in principle} means that we do not necessarily know which observable should be
measured, but the theory ensures that such exists. We shall see later that if the state of an {\em isolated}
system $I$\cite{f7}
is known, and this isolated system 
contains $R$, then the appropriate observable is $\hat \rho_R(I)$\cite{f9}. Therefore,
supposing that such an isolated environment always exist, the above reality criterion ensures
that all the states $\hat \rho_S(R_1)$, $\hat \rho_S(R_2)$... are elements of the reality.

The meaning of the quantum reference systems is now analogous
to the classical coordinate systems. Choosing a
classical coordinate system means that we imagine
what we would experience if we were there. Similarly,
choosing a quantum reference system $R$ means that we
imagine what we would experience if we did a nondisturbing measurement on
$R$ that does not disturbe $\hat \rho_R(R)=|\psi_R><\psi_R|$ 
while we learn it with probability $1$. Certainly, at the same time 
we also get $\hat \rho_S(R)=Tr_{R\setminus S} |\psi_R><\psi_R|$ for any $S\subset R$.

It is important to emphasize a specific new feature of the present approach: 
the above reality criterion {\em does not} imply that the states 
$\hat \rho_S(R_1)$, $\hat \rho_S(R_2)$... can all be {\em simultaneously} determined
via 
nondisturbing measurements. The reality criterion only ensures that {\em any} of them can be determined,
but usually only one (it may be freely choosen, which one) at the same time. The reason is very simple: a nondisturbing 
measurement done on the system $R_1$ does not change the state $\hat \rho_S(R_1)$, but it may
change the state $\hat \rho_S(R_2)$. Therefore, the fact that both $\hat \rho_S(R_1)$ and $\hat \rho_S(R_2)$
are elements of the reality does not imply that they can be compared. It is instructive to express
this feature (cf. the similar discussion in the Introduction) 
by telling that comparability of two states requires that not only the two states,
but also the {\em set} containing these two states be an {\em element} of reality. The fact that
each of the two states is an element of the reality implies only that the set containing the two states
is a {\em subset} of the reality.

So when can two (or more) states be compared? Consider the states $\hat \rho_{S_1}(R_1)$, $\hat \rho_{S_2}(R_2)$...
(which describe different systems if $S_1\ne S_2$). If the quantum reference systems
$R_1$, $R_2$ ... are {\em disjointed} (have no common subsystem) then the nondisturbing measurements
performed on $R_1$, $R_2$ ... do not disturbe each other, so the states can be compared, or, in other
terms, the set containing them is an element of the reality. Actually this is the only situation when
states are always comparable without changing them. 

\subsection{The new postulates}

As the dependence of $\hat \rho_S(R)$ on $R$
is a fundamental property now, one has to specify 
the relation of the different states in  terms of suitable
postulates. Although the motivation will be explained, the actual
justification of these postulates must come {\em a posteriori},
when checking their consistency and measurable consequences.

There will be two kinds of postulates. The first group ({\bf Postulates 2} and {\bf 5}) contains those
which have direct counterparts in traditional quantum mechanics and therefore
will seem mere translations to a new language. 
The second group ({\bf Postulates 1}, {\bf 3} and {\bf 4}), 
however, has no counterpart in traditional quantum mechanics.
These postulates render possible to free the theory from the {\em a priori}
classical background which was an essential part of the traditional theory.
Note that the present set of postulates is reduced compared to that given 
in Ref. \cite{bene}, in order to make the scheme more transparent. The two schemes
are completely identical, but in the present paper such rather obvious
statements like 'the state $\hat \rho_S(R)$ is a hermitian, positive
definite operator acting on the Hilbert space of $S$' have not
received the status of a postulate. 

{\bf Postulate 1. \em The internal state $\hat \rho_R(R)$ is always a single
dyad \hfill\break $|\Psi_R><\Psi_R|$.}

Therefore, we shall call the state $|\Psi_R>$ an internal state, too.

{\bf Postulate 2.} $\hat \rho_S(R)=Tr_{R\setminus S} |\psi_R><\psi_R|$

Here we have to add that despite of the different letters $R$ and $S$, 
there are no restrictions which physical system is the quantum reference system $R$
and which is the system $S$ to be described, except that $S\subseteq R$.

{\bf Postulate 3.\em  If the reference system $R=I$ is an isolated one
\cite{f7}
then the state $\hat \rho_S(I)$ commutes with the
internal state $\hat \rho_S(S)$.}

This means that the internal state of $S$ coincides with
one of the eigenstates of $\hat \rho_S(I)$. Knowing only $\hat \rho_S(I)$, 
it cannot be told, with which one. Only the corresponding probabilities can be given
(cf. below, {\bf Postulate 5}).  Therefore, we shall call 
the eigenstates of
$\hat \rho_S(I)$ the possible internal states of $S$
provided that the reference system $I$ is an
 isolated one. Let us emphasize that the relation between $\hat \rho_S(S)$ and $\hat \rho_S(I)$
 is many-to-one, and this makes possible to incorporate the indeterministic nature
 of quantum mechanics in the present approach.
 
 A frequent question is what happens if two (or more) nonzero eigenvalues\cite{f11}
 of $\hat \rho_S(I)$ coincide. This leads to a problem only if the eigenvalues are exactly
 equal. Certainly, if the eigenvalues are constant ($S$ does not interact with its environment, which is hardly
 realizable),
 this is a nonphysical situation, as two different physical quantities are never exactly equal.
 If $S$ interacts with its environment, the eigenvalues change and at a given instant of time $t_0$ they may
 cross each other. But this happens only at that instant of time, and not during some finite time interval.
 So the eigenstates are well defined as the function of time $t$, except at $t=t_0$.
 Taking the limit $\lim_{t\rightarrow t_0}$ of the eigenstates as functions of time, the limiting states
 may in principle define the eigenstates also at $t=t_0$, although a single instant of
 time is physically irrelevant.
  Certainly, if the eigenvalues are close to each other,
 the eigenstates become rather sensitive with respect to perturbations, but the exact degeneracy
 (when the eigenstates are truly undetermined) is a nonphysical case.\cite{f12} 
 
 Another question is whether the definition of $\hat \rho_S(I)$ is unique. Suppose $I_1$ and
 $I_2$ are two different isolated systems, both containing $S$. Be $R=I_1\cap I_2$. 
 Then, due to the definition of the isolated system, $I_1\setminus R$ has not yet interacted with
 $I_2$. As $I_2\supseteq R$, it also implies that $I_1\setminus R$ has not yet interacted with
 $R$. Therefore, $|\Psi_{I_1}>$ (the internal state of $I_1$)
 is of the product form $|\Psi_R>|\Psi_{I_1\setminus R}>$. Similarly,  $|\Psi_{I_2}>=|\Psi_R>|\Psi_{I_2\setminus R}>$.
 Therefore, $\hat \rho_S(I_1)=\hat \rho_S(I_2)=\hat \rho_S(R)$. Actually this uniqueness allows
 one to relate $\hat \rho_S(I)$ and $\hat \rho_S(S)$.

{\bf Postulate 4.\em  The result of a measurement is contained 
unambigously in the internal state of the measuring device.}

This is the only postulate which directly relates the present scheme to the experience.
One naturally expects that the state of the measuring device $M$ (as calculated using the rules
of the theory) contains the result of the measurement. But in the present theory
a given system has a multitude of states, each referring to a different quantum reference system.
One has to specify, which of them corresponds to the experience when doing a measurement. According to the
present postulate, it is the internal state $\hat \rho_M(M)$. If we return to the 
spin measurement, we find that this postulate (together with {\bf Postulate 3}) implies
that the measurement result corresponds either to the state $|m_\uparrow>$ or to the
state $|m_\downarrow>$. 

One may ask what is then a 'measuring device'. It can be either an observer (a person, you or me)\cite{f13}, or any system whose
internal states can be uniquely related to those of an observer. To be more explicit, suppose that one looks at
the measuring device in the spin measurement. The dynamics of this process can be 
written down schematically as
\begin{eqnarray}
|m_\uparrow>|O_0>\,\rightarrow\,|m_\uparrow>|O_\uparrow>\mbox{\hspace{2mm}}\nonumber\\
|m_\downarrow>|O_0>\,\rightarrow\,|m_\downarrow>|O_\downarrow>\,.\label{readoff}
\end{eqnarray}

Here e.g. $|O_\uparrow>$ stands for the state which describes that the observer
reads 'spin up' from the measuring device.
This {\em dynamics} is what qualifies the measuring device as being such, i.e., being able 
to convey information to the observer.
Using the postulates (including the last one, see below) we get that the internal state of the
measuring device and the observer are uniquely related. (This  will be shown in more detail
in the next subsection.) This unique relation depends on the special dynamics. If 
the possible internal states of the measuring device are such that the coordinates and momenta
of the pointer are well determined (i.e., they have small standard deviations), this special dynamics is realized
to a good approximation. This is actually the classical behavior. It should be emphasized
that classical behaviour in the present scheme is not something to be postulated. If
the theory is correct, it should come out automatically after having made realistic assumptions 
concerning the initial states, the structure and interactions of the systems we consider. This problem,
however, will not be discussed in the present paper.

{\bf Postulate 5. \em If there are $n$ ($n=1,\;2,\;3,\;...$) disjointed physical systems, 
denoted by
\hfill\break
$S_1, S_2, ... S_n$, all contained in the isolated reference 
system $I$ and 
having the 
possible internal states
$|\phi_{S_1,j}>,...,|\phi_{S_n,j}>$, respectively, 
then the joint
probability that $|\phi_{S_i,j_i}>$ 
coincides with the internal state of $S_i$ ($i=1,..n$)
is given by
\begin{eqnarray}
P(S_1,j_1,...,S_n,j_n)
=Tr_{S_1+...+S_n} [\hat \pi_{S_1,j_1} 
...\hat \pi_{S_n,j_n}\hat \rho_{S_1+...+S_n}(I)],\label{u5}
\end{eqnarray}
where $\hat \pi_{S_i,j_i}=|\phi_{S_i,j_i}><\phi_{S_i,j_i}|$.}

The possible internal states $|\phi_{S_i,j}>$ (cf. the definition after {\bf Postulate 3}) are
the eigenstates of $\hat \rho_{S_i}(I)$. It should be emphasized that this postulate is about
what exists (independent of measurement) and not about probabilities of the results of
measurements (as in the conventional interpretation). 

Let us mention briefly the special case $n=1$. The probability that 
an eigenstate of $\hat \rho_{S}(I)$ coincides
with the internal state of $S$ is given by the corresponding eigenvalue.

Finally, let us emphasize again that the time dependent Schr\"odinger equation remains valid for closed systems.
Note that in the present theory there is no collapse or reduction of the wave function,
and every conclusion should be drawn by using the above rules.

\subsection{Quantum measurements}

We are now in the position to discuss the problem of quantum measurements by using the new scheme.
Let us return to the spin measurement introduced at the beginning of subsection 2.1.
The internal state of the system $P+M$ (particle plus measuring device) after the
measurement is given by $|\Psi>$ in Eq. (\ref{u2}). If we want to know what the
result of the measurement is, then, according to {\bf Postulate 4} we have to determine
the internal state of the measuring device, i.e., $\hat \rho_M(M)$ . In order to do this first we
calculate the state of the measuring device with respect to the whole system $P+M$, i.e., 
we calculate $\hat \rho_M(P+M)$. According to {\bf Postulate 2} this is equal to
${\rm Tr}_P |\Psi><\Psi|=|m_{\uparrow}>|\alpha|^2<m_{\uparrow}|
+ |m_{\downarrow}>|\beta|^2<m_{\downarrow}|$ (cf. Eq. (\ref{u3}). 
Assuming that $P+M$ is isolated, {\bf Postulate 3} implies that
the internal state of $M$ will be one of the eigenstates of 
$\hat \rho_M(P+M)$, i.e., either $|m_{\uparrow}>$ or $|m_{\downarrow}>$.
In other terms, $\hat \rho_M(M)$ is either $|m_{\uparrow}><m_{\uparrow}|$ or $|m_{\downarrow}><m_{\downarrow}|$
(cf. {\bf Postulate 1}). 
From the knowledge of $\hat \rho_M(P+M)$ we cannot predict, which of these two state will $\hat \rho_M(M)$
actually be. We can, however, predict the corresponding probabilities: according to {\bf Postulate 5}
(applying it with $n=1$) $\hat \rho_M(M)=|m_{\uparrow}><m_{\uparrow}|$ with probability $|\alpha|^2$
and \hfill\break $\hat \rho_M(M)=|m_{\downarrow}><m_{\downarrow}|$ with probability $|\beta|^2$. 

One might say that this is just the same as in traditional quantum mechanics. Yes, it is.
But remember what was the price for it in traditional quantum mechanics: one had to introduce
the collapse of the wave function, i.e., one had to assume that the Schr\"odinger equation
is sometimes invalid (but it was not clearly specified, under which circumstances). The resulting
confusion is well demonstrated by Schr\"odinger's cat paradox. In contrast, in the present interpretation
the Schr\"odinger equation remains valid (it even describes the measurement process), and 
the correct result comes out when the postulates of the theory are consistently applied. 
It has also been shown in Ref. \cite{bene} that Schr\"odinger's cat paradox \cite{cat} can be solved
within the present approach.

Let us consider now the situation when an observer reads the result of the measurement. The dynamics of this
process is given approximately by Eq. (\ref{readoff}). Combining it with Eq. (\ref{u2}) we get for the
evolution of the internal state of the (by assumption isolated) $P+M+O$ system ($O$ standing for the observer)
during reading the device
\begin{eqnarray}
\left(\alpha |\uparrow>|m_{\uparrow}>
+\beta |\downarrow>|m_{\downarrow}>\right)|O_0>\mbox{\hspace{6cm}}\nonumber\\
\rightarrow\,
\alpha |\uparrow>|m_{\uparrow}>|O_{\uparrow}>
+\beta |\downarrow>|m_{\downarrow}>|O_{\downarrow}>\quad.
\end{eqnarray}
Applying {\bf Postulate 2}, we get 
\begin{eqnarray}
\hat \rho_P(P+M+O)=|{\uparrow}>|\alpha|^2<{\uparrow}|
+ |{\downarrow}>|\beta|^2<{\downarrow}|
\end{eqnarray}
and
\begin{eqnarray}
\hat \rho_M(P+M+O)=|m_{\uparrow}>|\alpha|^2<m_{\uparrow}|
+ |m_{\downarrow}>|\beta|^2<m_{\downarrow}|
\end{eqnarray}
both before and after the reading, while
\begin{eqnarray}
\hat \rho_O(P+M+O)=|O_{\uparrow}>|\alpha|^2<O_{\uparrow}|
+ |O_{\downarrow}>|\beta|^2<O_{\downarrow}|
\end{eqnarray}
after the reading. 
Applying {\bf Postulate 3} and {\bf Postulate 5} (for $n=1$) we get that 
$\hat \rho_P(P)=|{\uparrow}><{\uparrow}|$ with probability $|\alpha|^2$ or
$\hat \rho_P(P)=|{\downarrow}><{\downarrow}|$ with probability $|\beta|^2$,
$\hat \rho_M(M)=|m_{\uparrow}><m_{\uparrow}|$ with probability $|\alpha|^2$ or
$\hat \rho_M(M)=|m_{\downarrow}><m_{\downarrow}|$ with probability $|\beta|^2$.
These are valid both before and after the reading.
Similarly, we also get that after the reading
$\hat \rho_O(O)=|O_{\uparrow}><O_{\uparrow}|$ with probability $|\alpha|^2$ or
$\hat \rho_O(O)=|O_{\downarrow}><O_{\downarrow}|$ with probability $|\beta|^2$ .
Let us consider now the correlations between the internal states of the systems $P$ and $M$. 
Applying {\bf Postulate 5} for $n=2$ we find that the probability that
$\hat \rho_P(P)=|{\uparrow}><{\uparrow}|$ and $\hat \rho_M(M)=|m_{\downarrow}><m_{\downarrow}|$ is zero.
The probability that $\hat \rho_P(P)=|{\downarrow}><{\downarrow}|$ and $\hat \rho_M(M)=|m_{\uparrow}><m_{\uparrow}|$
is also zero. These results hold both before and after the reading. 

What does this all mean? First of all, it implies that
there is a strict correlation between the internal state of the measuring device
and that of the measured object. This is what makes us possible to draw conclusions about
the internal state of the measured object after the measurement from the knowledge of the 
measurement result. Secondly, in the above description the particle $P$ is a closed system
during the reading, which means an interaction between $M$ and $O$ only.
As has been emphasized above, for closed systems the Schr\"odinger equation holds.
In our case this implies that the internal state of the particle does not change when
reading the measuring device $M$. Moreover, as the correlation between the internal states 
of $P$ and $M$ is the same strict correlation both before and after the reading, it also
follows that that the internal state of the measuring device does not change when reading
the result. This again fits our physical expectations. 

Let us consider now the correlations between the internal states of the systems $P$ and $O$ and 
between those of $M$ and $O$. Applying {\bf Postulate 5} we find again a strict correlation.
E.g., if $\hat \rho_O(O)=|O_{\downarrow}><O_{\downarrow}|$, then 
$\hat \rho_P(P)=|{\downarrow}><{\downarrow}|$ and $\hat \rho_M(M)=|m_{\downarrow}><m_{\downarrow}|$.
Applying {\bf Postulate 4}, we can tell that the internal state of the observer corresponds
to that what he/she has read from the measuring device. The above strict correlations mean
that the knowledge of the measurement result allows one to draw unambigous conclusions about the internal states
of the measured object $P$ and of the measuring device $M$. Also, as the internal state of $M$ has not been
changed when reading the result, one may equally well apply {\bf Postulate 4} to $M$. This
demonstrates that in the present theory there is no ambiguity concerning the question when the measurement
is accomplished\cite{f14}. It also demonstrates that it is not necessary to calculate at each measurement situation
the internal state of the observer. 
It suffices to consider the measuring device only, provided that
 it is guaranteed that the possible internal states of the measuring device
will be uniquely related to those of an observer when reading the result. The classical nature of 
the possible internal states of the measuring device\cite{f15} 
ensures this.

Note that throughout the above discussion we used a rather simplified description. The more correct treatment
would have been the following. One considers the system $P+M$ together with all those systems which interact with it,
so that the resulting compound system $I\supset (P+M)$ is already (to a good approximation\cite{f16}) an isolated system. 
Provided the internal state of this system is known before the measurement, one
solves the time dependent Schr\"odinger equation to get the internal state of $I$ after the measurement
(remember that the term 'measurement' now means just the interaction between $P$ and $M$). Once it is known,
one applies {\bf Postulate 2} to get $\hat \rho_M(I)$ (i.e., one calculates ${\rm Tr}_{I\setminus M} \hat \rho_I(I)$). 
Applying now
{\bf Postulate 3} one has to diagonalize  $\hat \rho_M(I)$, as the internal state $\hat \rho_M(M)$
will be one of the eigenstates, and (due to {\bf Postulate 5}) 
an eigenvalue gives the probability that the corresponding eigenstate coincides 
with the internal state of $M$. Finally, according to {\bf Postulate 4} the internal state
of $M$ contains the result of the measurement. One can imagine this so that each eigenstate of $\hat \rho_M(I)$
is such that in coordinate representation it is sharply peaked at some definite pointer position
and in momentum representation it is sharply peaked at zero pointer momentum. 
This ensures that reading or recording the measurement result with a further device will not 
change the result of the measurement.

Let us return now to the question of the existence of the states. It has been told above
that the state $\hat \rho_S(R)$ of a system $S$ with respect to the quantum reference system
$R$ ($S\subseteq R$) is an element of the reality because in principle there exist a nondisturbing
measurement to it, i.e., if a suitable observable is measured on $R$ the state $\hat \rho_S(R)$
does not change but will be learned with probability $1$. We shall show this now.
Let us consider an isolated system $I$
which contains the system $R$. Consider the state $\hat \rho_R(I)$. This is at the same time a
hermitian operator, so it can be considered as an observable acting on the Hilbert space of $R$.
Suppose we are outside of $I$ and construct a measuring device $M$ (so that $M$ and $I$ are disjointed), 
whose interaction with $R$ can be given
by the relations
\begin{eqnarray}
|\phi_{R,j}>|m_0>\,\rightarrow\,|\phi_{R,j}>|m_j>\,\label{dyn}
\end{eqnarray}
where the states $|\phi_{R,j}>$ stand for the eigenstates of $\hat \rho_R(I)$, and the states $|m_j>$ are
the (orthonormed) pointer states of the measuring device.
The initial internal state of $I$ can be written as
\begin{eqnarray}
|\Psi_I>=\sum_j c_j |\phi_{I\setminus R,j}>|\phi_{R,j}>\,.\label{sch}
\end{eqnarray}
Here $|\phi_{I\setminus R,j}>$ stand for the eigenstates of $\hat \rho_{I\setminus R}(I)$.
Note that the expression (\ref{sch}) (Schmidt representation\cite{Schmidt}, see also Appendix A in Ref. 
\cite{bene}) is a completely general one. Due to the
properties of the orthonormed states $|\phi_{R,j}>$ and $|\phi_{I\setminus R,j}>$ 
only a single sum occurs in (\ref{sch}).
Combining (\ref{dyn}) and (\ref{sch}), we get for the state of the system $M+I$ after the measurement\cite{f17}
\begin{eqnarray}
\sum_j c_j |\phi_{I\setminus R,j}>|\phi_{R,j}>|m_j>\,.
\end{eqnarray}
Now we can repeat the calculation and the arguments we have had above when discussing 
the reading of the measuring device. Formally, the replacement of the systems is $P\rightarrow I\setminus R$,
$M\rightarrow R$, and $O\rightarrow M$. The conclusion now is that the internal state of $M$
is uniquely related to the internal state of $R$, and that this latter has not been changed
during the measurement. Therefore, if we get the $j$-th result when performing the measurement,
we can be sure that the internal state of $R$ is $\hat \rho_R(R)=|\phi_{R,j}><\phi_{R,j}|$,
and it has been the same already before the measurement. The state $\hat \rho_S(R)$ itself is given by
{\bf Postulate 2}.

\section{Explanation of the violation of Bell's inequality}

Let us consider now a two-particle system $P_1+P_2$ consisting of
the spin-half particles $P_1$ and $P_2$. They have previously interacted with each other
but then have become separated so much that there is no interaction between them
any longer. The initial
internal state of the two-particle system be
\begin{eqnarray}
\sum_j c_j
|\phi_{P_1,j}>|\phi_{P_2,j}>\label{u14}
\end{eqnarray}
where $c_1=a$, $c_2=-b$ (certainly $|a|^2+|b|^2=1$)\cite{f18}, $
|\phi_{P_1,1}>=|1,\uparrow>$, $
|\phi_{P_1,2}>=|1,\downarrow>$, $
|\phi_{P_2,1}>=|2,\downarrow>$, $
|\phi_{P_2,2}>=|2,\uparrow>$.
When the two-particle system is in the state (\ref{u14}),
there is a strict correlation
between the internal states $\hat \rho_{P_1}(P_1)$\hfill\break $=|\psi_{P_1}><\psi_{P_1}|$ 
and $\hat \rho_{P_2}(P_2)=|\psi_{P_2}><\psi_{P_2}|$. 
Provided that the system $P_1+P_2$
is initially isolated, applying {\bf Postulate 5} we obtain that the
probability that $|\psi_{P_1}>=|\phi_{P_1,j}>$ and 
$|\psi_{P_2}>=|\phi_{P_2,k}>$ is
$
P(P_1,j,P_2,k)=|c_j|^2\delta_{j,k}
$.

Let us consider now a typical experimental situation, 
when measurements on both
particles are performed. Before the measurements the internal
state of the isolated system $P_1+M_1+P_2+M_2$ ($P_1,P_2$
standing for the particles and $M_1,M_2$ for the measuring
devices, respectively) is given by
$
\left(\sum_j c_j
|\phi_{P_1,j}>|\phi_{P_2,j}>\right)|m^{(1)}_0> |m^{(2)}_0>$,
while it is
\begin{eqnarray}
\sum_j c_j
\hat U_t(P_1+M_1)\left(|\phi_{P_1,j}>|m^{(1)}_0>\right)
\,\hat U_t(P_2+M_2)\left(|\phi_{P_2,j}>|m^{(2)}_0>\right) \quad,\label{u15}
\end{eqnarray}
with a time $t$ later, i.e. during and after the measurements. Here 
$\hat U_t(P_i+M_i)$ ($i=1,2$) stands for the unitary time evolution operator
of the closed system $P_i+M_i$.

Eq.(\ref{u15}) implies (according to {\bf Postulate 2}) that
\begin{eqnarray}
\hat \rho_{P_1+M_1}(P_1+M_1+P_2+M_2)\mbox{\hspace{7.5cm}}\\=
\sum_j \hat U_t(P_1+M_1)|\phi_{P_1,j}>|m^{(1)}_0>
|c_j|^2 <m^{(1)}_0|<\phi_{P_1,j}|\hat U^+_t(P_1+M_1)\nonumber
\end{eqnarray}
and
\begin{eqnarray}
\hat \rho_{P_2+M_2}(P_1+M_1+P_2+M_2)\mbox{\hspace{7.5cm}}\\
=\sum_j \hat U_t(P_2+M_2)|\phi_{P_2,j}>|m^{(2)}_0>
|c_j|^2 <m^{(2)}_0|<\phi_{P_2,j}|\hat U^+_t(P_2+M_2)\,.\nonumber
\end{eqnarray}
Here we have made used the unitarity of the operators $\hat U_t(P_i+M_i)$ when
calculating the traces.
According to {\bf Postulate 3} the internal state of 
the $P_1+M_1$ is one of the states $\hat U_t(P_1+M_1)|\phi_{P_1,j}>|m^{(1)}_0>$ and 
that of $P_2+M_2$ is one of the states $\hat U_t(P_2+M_2)|\phi_{P_2,j}>|m^{(2)}_0>$, the 
corresponding probabilities being in both cases (according to {\bf Postulate 5}) $|c_j|^2$.
One can see here explicitly that the possible internal states of $P_1+M_1$ are independent
of the interaction between $P_2$ and $M_2$, and vice versa. 
As the systems $P_1+M_1$ and $P_2+M_2$ are closed systems, their internal states evolve
unitarily which also means that if initially (at time $t=0$) the internal state
of $P_1+M_1$ was $|\phi_{P_1,1}>|m^{(1)}_0>$, then at a later time $t$
it is $\hat U_t(P_1+M_1)|\phi_{P_1,1}>|m^{(1)}_0>$. 
Hence we see that according to the present theory the two measurement processes, 
i.e., the time evolution of the internal state of $P_1+M_1$  and of  $P_2+M_2$,
do not influence each other. This means that there is no mysterious action-at-a-distance,
and the principle of locality is satisfied. But why is then Bell's inequality violated?
This will be explained below. Before doing that, it is of worth addressing here briefly one more 
question. One may ask what is then the state of $P_1+M_1$ one gets if one applies the traditional
concept of the collapse of the wave function to the second measurement process.
That state {\em does} depend on the type and result of the second measurement.
It turns out that it can be identified with  $\hat \rho_{P_1+M_1}(P_1+M_1+P_2)$ 
in the present scheme (remember that
the internal state is $\hat \rho_{P_1+M_1}(P_1+M_1)$). It can be understood why the state 
$\hat \rho_{P_1+M_1}(P_1+M_1+P_2)$ depends on the second measurement. It is not because of
some action-at-a-distance between $P_1+M_1$ and $P_2+M_2$, 
but it is due to the change of the internal state of the quantum reference
system $P_1+M_1+P_2$. This change is indeed due to the second measurement process,
but takes place locally, as the second particle which interacts with $M_2$ is a subsystem of $P_1+M_1+P_2$.
This is actually the resolution of the EPR paradox within the framework of the present approach\cite{bene}.

The time evolution of the internal state of $P_i+M_i$ can be given more explicitly 
if we specify the dynamics of the measurements by the 
relations
\begin{eqnarray}
|\xi(P_i,j)>|m^{(i)}_0>\;\rightarrow \;|\xi(P_i,j)>|m^{(i)}_j>\quad,
\label{u16}
\end{eqnarray}
where $i,j=1,2$, $|\xi(P_1,j)>$ is the $j$-th eigenstate of the 
spin component $\vec{S} \cdot \vec{a}$ measured 
on the first particle and $|\xi(P_2,j)>$ is the $j$-th eigenstate of the 
spin component $\vec{S} \cdot \vec{b}$ measured 
on the second particle (cf. the notations in the Introduction). Using Eq.(\ref{u16}) 
the time evolution of 
the internal state of the closed systems $P_i+M_i$ is given explicitly by
\begin{eqnarray}
|\psi_{P_i}>|m_0^{(i)}>\;\rightarrow \;
\sum_j <\xi(P_i,j)|\psi_{P_i}>|\xi(P_i,j)>|m_j^{(i)}>\,.\label{xx1}
\end{eqnarray}
One can see again that the $i$-th measurement process 
is completely determined by the initial internal state of the
particle $P_i$. Therefore, any correlation between
the measurements may only stem from the initial correlation
of the internal states of the particles.

Let us calculate now the internal state $\hat \rho_{M_1}(M_1)$ which corresponds to the
measured value $a$ (cf. {\bf Postulate 4}). 
Using Eq.(\ref{u16}) the final internal state (\ref{u15}) of the whole isolated system
$P_1+P_2+M_1+M_2$ may be written as
\begin{eqnarray}
\sum_{j,k}\left(
\sum_l c_l<\xi(P_1,j)|\phi_{P_1,l}><\xi(P_2,k)|\phi_{P_2,l}>
\right)\nonumber\\
\times |\xi(P_1,j)>|\xi(P_2,k)>|m^{(1)}_j>|m^{(2)}_k>\,.
\end{eqnarray}
According to {\bf Postulate 2} $\hat \rho_{M_1}(P_1+P_2+M_1+M_2)$
can be expressed as 
\begin{eqnarray}
\hat \rho_{M_1}(P_1+P_2+M_1+M_2)\mbox{\hspace{5cm}}\nonumber\\
=\sum_j\left(\sum_l
|c_l|^2 |<\xi(P_1,j)|\phi_{P_1,l}>|^2\right)|m^{(1)}_j><m^{(1)}_j|\,.
\end{eqnarray}
Note that it is independent of the second measurement. 
According to {\bf Postulate 3}  the internal state of $M_1$ is one of the
$|m^{(1)}_j>$-s. 
The probability to observe the $j$-th result (up or
down spin in the direction $\vec{a}$) is (according to {\bf Postulate 5})
\begin{eqnarray}
P(M_1,j)=\sum_l |c_l|^2 |<\xi(P_1,j)|\phi_{P_1,l}>|^2\,.
\end{eqnarray}
This may be interpreted as one intuitively expects: $|c_l|^2$ is
the probability that the initial internal state $|\psi_{P_1}>$ is  just $|\phi_{P_1,l}>$ (cf. {\bf Postulate 5}),
and $|<\xi(P_1,j)|\phi_{P_1,l}>|^2$ is the conditional
probability that one gets the $j$-th result if $|\psi_{P_1}>=|\phi_{P_1,l}>$.

Let us show that the latter statement also follows from our postulates.
Applying {\bf Postulate 5} with $n=2$ we get that the internal states of $P_1$ and
$P_2$ are uniquely related before the measurements. Moreover,
the initial internal state of $P_2$ uniquely determines the final internal state
of the system $P_2+M_2$ (this follows in a completely analogous way as in case
of the first measurement process discussed above). Hence, the internal state
of the system $P_2+M_2$ after the measurements is still uniquely related to the
initial internal state of $P_1$. Therefore, the joint probability
that initially the internal state of $P_1$ coincides with $|\phi_{P_1,l}>$ and
the result of the first measurement (represented by the internal state of $M_1$,
cf. {\bf Postulate 4}) corresponds to $|m^{(1)}_j>$ is the same as the 
 joint probability that after the measurement the internal state of
$P_2+M_2$ coindides with its $l$-th possible internal state 
$$\sum_k <\xi(P_2,k)|\phi_{P_2,l}>|\xi(P_2,k)>|m_k^{(2)}>$$ 
(cf. Eq. (\ref{xx1}))  and
the internal state of $M_1$ coindides with its $j$-th possible internal state $|m^{(1)}_j>$.
In case of this latter probability the systems involved, i.e. $M_1$ and $P_2+M_2$ are disjointed,
thus we may apply {\bf Postulate 5} to get the expression
$|c_l|^2 |<\xi(P_1,j)|\phi_{P_1,l}>|^2$ which already implies the above interpretation
of the factors.

Certainly the first and the second measurement play a symmetric role.
Thus we have shown that the initial internal state of $P_1$ ($P_2$) determines
the outcome of the first (second) measurement in the usual 
probabilistic sense. 

But doesn't it mean that the internal states of $P_1$ and $P_2$
are local hidden variables? They indeed play the role of $\lambda$ (cf. the Introduction).
Nevertheless, there is a fundamental difference. Hidden variables
are thought to be comparable with the results of the measurements
so that their joint probability $P(a,b,\lambda)$ can be defined (cf. assumption (*) in the
Introduction).  In contrast, we show below that in the present theory
there is no way to define the analogous joint probability
$P(P_1,l_1,P_2,l_2,(0);M_1,j,M_2,k,(t))$, i.e., the probability that initially
the internal state of $P_1$ is $|\phi_{P_1,l_1}>$ and that of $P_2$ is $|\phi_{P_2,l_2}>$
{\em and} finally the internal state of $M_1$ is $|m^{(1)}_j>$ 
and that of $M_2$ is $|m^{(2)}_k>$. Intuitively we would write
\begin{eqnarray}
P(P_1,l_1,P_2,l_2,(0);M_1,j,M_2,k,(t))\mbox{\hspace{3.5cm}}\nonumber\\
=|c_{l_1}|^2\delta_{l_1,l_2}|<\xi(P_1,j)|\phi_{P_1,l_1}>|^2\,
<\xi(P_2,k)|\phi_{P_2,l_2}>|^2,\label{u18}
\end{eqnarray}
as $|c_{l_1}|^2\delta_{l_1,l_2}$ is
the joint probability that $|\psi_{P_1}>=|\phi_{P_1,l}>$ \cite{f19}
and $|\psi_{P_2}>=|\phi_{P_2,l}>$, and $|<\xi(P_i,j)|\phi_{P_i,l_i}>|^2$ is the conditional
probability that one gets the $j$-th result in the $i$-th
measurement if initially $|\psi_{P_i}>=|\phi_{P_i,l_i}>$ ($i=1,2$).
Certainly the existence of such a joint probability would immediately imply the
validity of Bell's inequality, thus it is absolutely important 
to understand why this probability does not exist.

Let us mention, first of all, that using  {\bf  Postulate 5} for $n=2$, we may calculate the correlation
between the measurements, i.e., the joint probability 
that $|\psi_{M_1}>=|m^{(1)}_j>$ {\em and} $|\psi_{M_2}>=|m^{(2)}_k>$.
We obtain
\begin{eqnarray}
P(M_1,j,M_2,k)
=\left|\sum_l c_l<\xi(P_1,j)|\phi_{P_1,l}><\xi(P_2,k)|\phi_{P_2,l}>\right|^2
\;.\label{u19}
\end{eqnarray}
This is the usual quantum mechanical expression 
which violates Bell's inequality and whose correctness is experimentally
proven. Thus our theory gives the correct expression for the correlation.
Nevertheless, if the joint probability 
(\ref{u18}) exists, it leads to 
\begin{eqnarray}
P(M_1,j,M_2,k)
=\sum_l |c_l|^2 |<\xi(P_1,j)|\phi_{P_1,l}>|^2
|<\xi(P_2,k)|\phi_{P_2,l}>|^2\label{u20}
\end{eqnarray}
which satisfies Bell's inequality and contradicts
Eq.(\ref{u19}). 
Let us demonstrate that no such contradiction appears.

If we try to find an expression for
the joint probability \hfill\break
$P(P_1,l_1,P_2,l_2,(0);M_1,j,M_2,k,(t))$ by using our Postulates,
the first difficuly is that the occuring internal states are given
at different times, while {\bf Postulate 5} (which gives the joint probabilities
in the present theory) refers only to equal time correlations. 
Nevertheless, we can use the fact that the initial internal state of
$P_1$ ($P_2$) is uniquely related to the final internal state of $P_1+M_1$ ($P_2+M_2$).
Therefore, if $P(P_1,l_1,P_2,l_2,(0);M_1,j,M_2,k,(t))$ exists,
then it must be equal to $P(P_1+M_1,l_1,P_2+M_2,l_2,M_1,j,M_2,k)$. 
In this latter joint probability all the possible internal states
are given at the final time (i.e., after the measurements). But 
still we cannot apply {\bf Postulate 5}, as the systems that occur here
are not disjointed. We may tell that the internal state of $P_1+M_1$ and that of $P_2+M_2$
are uniquely related, so it is enough to consider the joint probability
$P(P_1+M_1,l,M_1,j,M_2,k)$. But even here the systems $P_1+M_1$ and $M_1$ are
not disjointed. If we still try to apply Eq. (\ref{u5}), the result will depend on the
ordering of the projectors $\hat \pi_{P_1+M_1,l}=|\phi_{P_1+M_1,l}><\phi_{P_1+M_1,l}|$
and $\hat \pi_{M_1,j}=|\phi_{M_1,j}><\phi_{M_1,j}|$. We get either\cite{f20}
\footnotesize \begin{eqnarray}
P(P_1+M_1,l,M_1,j,M_2,k)\mbox{\hspace{8.7cm}}\nonumber\\
=Tr_{P_1+M_1+M_2} [\hat \pi_{P_1+M_1,l} 
\hat \pi_{M_1,j}\hat \pi_{M_2,k}\hat \rho_{P_1+M_1+M_2}(P_1+P_2+M_1+M_2)]\mbox{\hspace{2.2cm}}\label{xx2}\\
=c_l^*<\phi_{P_1,l}|\xi(P_1,j)><\phi_{P_2,l}|\xi(P_2,k)>
\sum_n c_n <\xi(P_1,j)|\phi_{P_1,n}><\xi(P_2,k)|\phi_{P_2,n}>\nonumber
\end{eqnarray}\normalsize
or 
\footnotesize \begin{eqnarray}
P(P_1+M_1,l,M_1,j,M_2,k)\mbox{\hspace{8.9cm}}\nonumber\\
=Tr_{P_1+M_1+M_2} [ 
\hat \pi_{M_1,j}\hat \pi_{P_1+M_1,l}\hat \pi_{M_2,k}\hat \rho_{P_1+M_1+M_2}(P_1+P_2+M_1+M_2)]\mbox{\hspace{2.4cm}}\label{xx3}\\
=c_l<\xi(P_1,j)|\phi_{P_1,l}><\xi(P_2,k)|\phi_{P_2,l}>
\sum_n c_n^* <\phi_{P_1,n}|\xi(P_1,j)><\phi_{P_2,n}|\xi(P_2,k)>\,.\nonumber
\end{eqnarray}\normalsize
These expressions are usually not even real\cite{f21} 
so one cannot interpret them as probabilities. Therefore, the joint probability\hfill\break
$P(P_1,l_1,P_2,l_2,(0);M_1,j,M_2,k,(t))$ cannot be defined within the framework of the present
theory.

One can arrive at the same conclusion in another, rather instructive way. 
Instead of trying to find a suitable formal expression, one may
try to define the joint probability in question in terms
of suitable measurements. Measurements should of course be
treated as in Section 3., i.e., they are interactions between
physical systems and obey the postulates of the theory.

Using the method proposed above, we shall always get real, nonnegative
expressions for the probabilities, as we shall consider correlations
among the disjointed measuring devices. Let us
 determine the initial internal state of $P_1$ by a suitable nondisturbing measurement,
so that we can compare the result with those of the other two
measurements later. The dynamics of the nondisturbing measurement is given
by 
\begin{eqnarray}
|\phi_{P_1,l}>|m_0^{(3)}>\,\rightarrow\,|\phi_{P_1,l}>|m_l^{(3)}>\,.
\end{eqnarray}
Here the states $|m_0^{(3)}>$, $|m_l^{(3)}>$ are states of the extra measuring device $M_3$.
Therefore, the internal state of the system $P_1+P_2+M_1+M_2+M_3$ after the extra, nondisturbing measurement,
but before the other two original measurements is given by
\begin{eqnarray}
\left(\sum_l c_l |\phi_{P_1,l}>|\phi_{P_2,l}>|m_l^{(3)}>\right)|m_0^{(1)}>|m_0^{(2)}>\,.
\end{eqnarray}
One can show (cf. the discussion at the end of Section 3.) that the internal state
of $P_1$ has not changed, but the internal state of the measuring device $M_3$
is uniquely related to it (and to the internal state of $P_2$ as well).
Using Eq. (\ref{u16}) one obtains for the final internal state of the system $P_1+P_2+M_1+M_2+M_3$
the expression
\begin{eqnarray}
\sum_l c_l\left( \sum_j <\xi(P_1,j)|\phi_{P_1,l}>|\xi(P_1,j)>|m^{(1)}_j>\right)\mbox{\hspace{1.5cm}}
\nonumber\\
\times\left( \sum_k <\xi(P_2,k)|\phi_{P_2,l}>|\xi(P_2,k)>|m^{(2)}_k>
\right)|m^{(3)}_l>\;.\label{u21}
\end{eqnarray} 
As the systems $M_1,\;M_2,\;M_3$ are disjointed,
we may apply {\bf Postulate 5} for $n=3$ and we indeed get for 
$P(M_3,l,M_1,j,M_2,k)$ the expression
(\ref{u18}) (for $l_1=l_2=l$). Do we get then a contradiction with Eq.(\ref{u19})?
No, because applying {\bf Postulate 5} directly for $n=2$, we get in this case
Eq.(\ref{u20}) instead of Eq.(\ref{u19}). Thus we see that the
extra measurement has changed the correlations and our theory 
gives account of this effect consistently. So the situation is that if we do not
perform the extra measurement with $M_3$, then the joint probability
$P(P_1,l_1,P_2,l_2,(0);M_1,j,M_2,k,(t))$ cannot be defined, therefore
the assumption (*) made at the derivation of Bell's inequality fails
and the correlation (\ref{u19}) violates Bell's inequality. If, however,
we do the extra measurement, $P(P_1,l_1,P_2,l_2,(0);M_1,j,M_2,k,(t))$
can be defined via $P(M_3,l,M_1,j,M_2,k)$, but the extra measurement 
changes the correlations $P(M_1,j,M_2,k)$, so that they indeed satisfy Bell's inequality.

Summing up, we have seen that the initial internal state
of $P_1$ ($P_2$) determines the first (second) measurement
process. As the measurement processes do not influence each other, 
the observed correlations may stem only from the 'common past'
of the particles. The internal state of $P_1$ and that of $P_2$
 'carry' the initial correlations
and 'transfer' them to the measuring devices. 
In this sense the internal state of $P_1$ (or that of $P_2$)
plays the role of the hidden parameter $\lambda$. This state, however,
is not hidden at all, as it can be determined via a suitable measurement
with unit probability, without disturbing this state itself.
Another difference is that in the absence of such a measurement, 
the initial internal state of $P_1$ is not comparable with the results of the
measurements done by $M_1$ and $M_2$, so that a joint probability to these
states cannot be assigned.
 
This means that the reason for the
 violation of Bell's inequality is that the usual derivations
 always assume that those quantities (hidden parameters)
 which carry the initial correlations can be freely compared with the results
 of the measurements. This comparability is usually 
 thought to be a consequence of realism.
 According to the present theory, the above assumption
 goes beyond the requirements of realism and proves to be wrong,
 as e.g. the internal states $|\psi_{P_1+M_1}>$, 
$|\psi_{M_1}>$ and $|\psi_{M_2}>$ do exist individually,
but they cannot be compared without changing the correlations.

On the other hand, any attempt to
compare the initial internal states of $P_1$ and $P_2$ with
the results of both measurements changes the correlations,
so that they satisfy Bell's inequality. The present theory gives account of this
effect consistently.

\section{Conclusion}
A new interpretation of quantum mechanics has been discussed.
It has been demonstrated that the consistent use of its postulates 
leads to a physically acceptable description of the quantum measurements.

Most importantly, the EPR-Bell experiment has been analyzed within the framework of the new
approach.
The main conclusion is that Bell's theorem (which states that no local realistic theory
can reproduce all the statistical correlations of quantum mechanical two-body systems) 
is {\em not valid}. Quantum mechanics
{\em is} a local realistic theory. The violation of Bell's inequality is due to
the fact that certain states, although exist, cannot be compared 
without changing the experimental correlations and therefore a corresponding joint 
probability cannot be defined. 

Note that there are other proofs of Bell's theorem which
do not use Bell's inequality (or any other kind of inequality)\cite{other}. Nevertheless,
these proofs also assume that existing quantities (including the 'hidden parameters', whose role
is played by some quantum states within the present approach) can always be compared, 
an assumption which usually does not hold true according to the present theory. 
It can in fact be shown that these proofs are not correct from the point of view of the new theory. 
The details of this, however, will be published elsewhere.

Finally, let me comment briefly the relation of the present interpretation to the previous ones.
There is an enormous number of different attempts, and there are quite a few which are
based on the assumption of the universal validity of the Schr\"odinger equation \cite{altern}, \cite{altern1},
\cite{altern2}.
Some even utilize the Schmidt decomposition, which is strongly related to Postulate 3.\cite{altern2}
The present interpretation possesses these features, too.
Nevertheless, it has some essentially new features as well. These are the following:
\begin{enumerate}
\item the concept of the quantum reference systems,
\item the assumption that the dependence of quantum states on quantum reference systems
is a fundamental property of quantum mechanics,
\item the consistent formulation of the rules of the theory,
\item the present interpretation implies that quantum mechanics is a local realistic theory.
\end{enumerate}
It should be noted that two previous interpretations have introduced some
rather similar (but not equivalent) concept as the 
the dependence of quantum states on quantum reference systems here. In Ref. \cite{altern}
states of a system are defined with respect to {\em the states of the complementer subsystem}.
In Ref. \cite{altern}
states of a system are defined with respect to {\em the complementer subsystems (observers)}.
In contrast, in the present interpretation states of a system are defined 
with respect to the {\em quantum reference system,
which contains the system to be described}. 
Note that the above interpretations, contrary to the present one,
do not explain the violation of Bell's inequality by maintaining realism and the principle of locality.

\vskip2cm

\centerline{\bf Acknowledgements}
\vskip0.5cm
The author is indebted to A.Bringer, G.Eilenberger, M.Eisele, T.Geszti,
R.Graham, G.Gy\"orgyi, F.Haake, Z.Kaufmann, H.Lustfeld, P.Rosenqvist,\hfill\break Y.Shih, 
P.Sz\'epfalusy, T.Tasn\'adi and G.Vattay for useful discussions,\hfill\break
to P.Sz\'epfalusy also for his continued interest in 
 the work and for encouragement, and to N.D.Mermin for several
 valuable remarks and advices.
 
 The author wants to thank for the hospitality of the {\em Institut f\"ur 
 Festk\"orper\-physik, Forschungszentrum J\"ulich GmbH} where a 
 substantial 
 part of the work has been done. 
 The present paper came into being 
 within the framework of a scientific and technological cooperation agreement
 between the Hungarian and the German government 
 as a result of a research cooperation supported
 by the OMFB (Hungary) and the BMFT (Germany).
 This work has been partially supported by the Hungarian Academy of 
 Sciences
 under Grant Nos. OTKA T 017493, OTKA F 17166 and OTKA F 19266.


\begin{thebibliography}{99}
\bibitem{f2} Note that Bell 
derived his inequality for the expectation value of the product of the
spins, i.e., for $E(\vec a,\vec b)=P(\vec a\,+,\vec b\,+)-P(\vec a\,+,\vec b\,-)
-P(\vec a\,-,\vec b\,+)+P(\vec a\,-,\vec b\,-)=4 P(\vec a\,+,\vec b\,+)
-2 P_1(\vec a\,+)-2 P_2(\vec b\,+)+1=4 P(\vec a\,+,\vec b\,+)-1$. Eq.(\ref{e3}) readily implies
\begin{eqnarray}
1+E(\vec a,\vec b)\ge \left|E(\vec a,\vec c)-E(\vec b,\vec c)\right|\;,\nonumber
\end{eqnarray}
which is the original form of Bell's inequality.
\bibitem{f3} Note that in this experiment
photons were used, therefore Eq.(\ref{e4}), which refers to spin-half
particles, does not apply.
\bibitem{f5} This is due to the absence
of a distinguished inertia system and the finiteness 
of the velocity of light in vacuum.
\bibitem{f6} Until the rules of the new theory are set,
we have to use the traditional framework.
\bibitem{f7} We define an isolated system as
such a system which has not been interacting 
with the outside world. In contrast, a closed system
is such a system that is not interacting with any other
system at the given instant of time 
(but might have interacted in the past).
\bibitem{f9} As
$\hat \rho_R(I)$ is a hermitian operator, formally it can be conceived as an observable as well.
\bibitem{f13} After all,
experience is human experience.
\bibitem{f14} As is known, in traditional quantum mechanics there is no clear answer to this
question. One may tell, the measurement is accomplished when the interaction between the
measuring device and the measured object has happened. But one may tell equally well that it is accomplished only
if the result has been read by an observer or even if we ourselves asked this observer about
the result. In contrast, in the new approach the measurement is unambigously 
accomplished when the interaction between the
measuring device and the measured object has happened.
\bibitem{f16} 'Good approximation'
means here that a further enlagement of the compound system does not lead to a significant difference in the
results of the procedure described below.
\bibitem{f17} Here
we tacitly assume that $M$ is an isolated system as well. Strictly speaking, we should have included in $M$
all the systems which interacted with it before the measurement (even ourselves, as we have interacted with
it during its construction) to have an actually isolated system. One can convince himself/herself, however, that
such a more careful description would give the same result.
\bibitem{f18} We assume that $|a|$ and $|b|$ are not exactly equal, as
it leads to the nonphysical degeneracy mentioned after Postulate 3.
\bibitem{f20} There are only two
different cases. This is due to the possibility of a cyclic rearrangements of the operators under the trace
and also because $\hat \pi_{M_2,k}$ commutes with the other two projectors.
\bibitem{f21} This is the formal
reason why one has to consider disjointed systems when defining joint probabilities.
If the systems are disjointed, expression (\ref{u5}) is real and nonnegative. 
If the systems are not disjointed, even reality cannot be satisfied. Note that Eq. (\ref{xx3}) is just
the complex conjugate of Eq. (\ref{xx2}). It is interesting to see that summing over $l$
(which corresponds to integrating over $\lambda$, cf. the Introduction) 
leads to the correct correlation (\ref{u19}) in both cases.
\end{thebibliography}
\end{document}